# Normalization at the field level: fractional counting of citations

Loet Leydesdorff [i] & Tobias Opthof [ii]

Van Raan *et al.* (2010) accepted our critique for the case of journal normalization (previously *CPP/JCSm*); CWTS has in the meantime adapted its procedures.[1] However, a new indicator was proposed for field normalization (previously *CPP/FCSm*), called the "mean normalized citation score" (*MNCS*; cf. Lundberg, 2007).[2] In our opinion, this latter change does not sufficiently resolve the problems. Since the new indicator is also considered another "crown indicator," it seems urgent to warn against and elaborate on these remaining problems. In addition to damaging evaluation processes at the level of individuals and institutions, the "crown indicator" is also used by CWTS for the "Leiden Rankings," and flaws in it can therefore misguide policies at national levels.

We focused in Opthof & Leydesdorff (2010) on journal normalization because in the case of field normalization, one has two problems: the scientometric one of how to delineate fields and the statistical one of how to normalize. Journal normalization is the simpler case because journals are delineated units of analysis. Like *CPP/FCSm*, *MNCS* is based on the ISI Subject Categories for weighing citation scores at the field level. The ISI Subject Categories, however, were not designed for the scientometric evaluation, but for the purpose of information retrieval. Despite a strong denial by Van Raan *et al.* (2010) who formulate: "we are not aware of any convincing evidence of large-scale inaccuracies in the classification scheme of WoS," the subject categories lack an analytical base (Pudovkin & Garfield, 2002, at p. 1113n.; Rafols & Leydesdorff, 2009) and are not literary-warranted (Bensman & Leydesdorff, 2009). Several alternatives for the classification have been proposed (Bornmann, 2010; Glänzel & Schubert, 2003).

In our opinion, the purpose of normalization at the field level is to control for differences in expected citation frequencies among fields (Garfield, 1979, at p. 366; McAllister *et al.*, 1983; Moed, 2010b). These differences are caused by differences in citation behavior among scholars in various fields of science. Mathematics, for example, is known to have a much lower citation density than the biomedical sciences. In our opinion, the easiest way to capture the differences in citation behavior among fields is by fractional counting in the citing articles at the article level (Small & Sweeney, 1985). For example, if an author in mathematics cites six references, each reference can be counted as 1/6 of overall citation, whereas a citation in a paper in biomedicine with 40 cited references can be counted as 1/40. This normalization thoroughly takes field differences into account and the results allow for statistical testing. Most importantly, the normalization is independent from a classification system and thus there is no indexer effect.

---

[1] Moed (2010a) argues for using the old *CPP/JCSm* and *CPP/FCSm* ratios because at the level of aggregates (groups or oeuvres) distributions are less important, in his opinion. However, one should distinguish between the aggregation of units of analysis and the normalization of variables. Distributions of variables over cases are crucial for testing the significance of observed differences, both at the level of individual cases and at the level of groups or their oeuvres.

[2] The *MNCS* indicator is not to be confused with the existing indicator *NMCR* or Normalized Mean Citation Rate used by ECOOM in Leuven (Glänzel *et al.*, 2009, at p. 182). The *NMCR* of ECOOM-Leuven is equivalent to the old "crown indicator" (*CPP/FCSm*) of CWTS-Leiden.

Let us thus turn this critique into a constructive proposal by showing the difference between the journal normalization contained in our previous contribution to this debate and the field normalization proposed here using the same seven PIs in our sample of the 232 scientists evaluated at the Academic Medical Center in Amsterdam, The Netherlands.

| | *Bibliometric data* | | | *Journal normalized* | | *Field normalized* | | |
|---|---|---|---|---|---|---|---|---|
| *Rank* | $\Sigma p_i$ | $\Sigma c_i$ | *Avg(c/p)* | Mean citation score (previous study) | *CPP/JCSm* (CWTS, 2008) | $\Sigma c_f$ (this study) | *Avg($c_f$)* (this study) | *CPP/FCSm* (CWTS, 2008) |
| 6 | 23 | 891 | 38.74 (± 13.67) | 2.03 (± 0.55) | 2.18 | 31.95 | 1.39 (± 0.50) | 2.94 |
| 14 | 37 | 962 | 26.00 (±  4.09) | 1.74 (± 0.19) | 1.86 | 30.32 | 0.82 (± 0.13) | 3.20 |
| 26 | 22 | 567 | 25.77 (±  5.78) | 1.54 (± 0.23) | 1.56 | 21.74 | 0.99 (± 0.25) | 2.17 |
| 117 | 32 | 197 | 6.16 (±  1.30) | 1.50 (± 0.29) | 1.00 | 6.83 | 0.21 (± 0.44) | 0.92 |
| 118 | 37 | 402 | 10.86 (±  2.21) | 0.93 (± 0.13) | 1.00 | 16.08 | 0.43 (± 0.09) | 1.43 |
| 206 | 65 | 647 | 9.96 (±  1.57) | 0.91 (± 0.11) | 0.58 | 21.90 | 0.34 (± 0.05) | 0.87 |
| 223 | 32 | 354 | 11.06 (±  1.74) | 0.78 (± 0.12) | 0.43 | 12.40 | 0.39 (± 0.08) | 0.72 |
| | | | | Spearman $\rho > 0.99$; $p < 0.01$ Pearson's $r = 0.94$; $p < 0.01$ | | | Spearman $\rho = 0.75$; *n.s.* Pearson's $r = 0.85$; $p < 0.05$ | |

**Table 1**: The effects of different normalizations on values and ranks

Table 1 shows the different normalizations. The journal normalizations in the middle of this table correspond to the figures provided in Table 4 of Opthof & Leydesdorff (2010). These *journal*-normalized ranking correlate highly in terms of their rank ordering (Spearman's $\rho > 0.99$; $p < 0.01$) but, as we argued, there are considerable differences at the level of individual scores. However, the two *field* normalizations—this study *versus* CWTS (2008)—correlate much less strongly. Given the strong correlations between the new and old "crown indicators" (Van Raan *et al.*, 2010, at p. 5), the new "crown indicator" can be expected to inherit the flaws of the old one. These problems are unnecessarily generated by using the ISI Subject Categories for the normalization at the field level (Leydesdorff & Opthof, in preparation; Rafols & Leydesdorff, 2009).

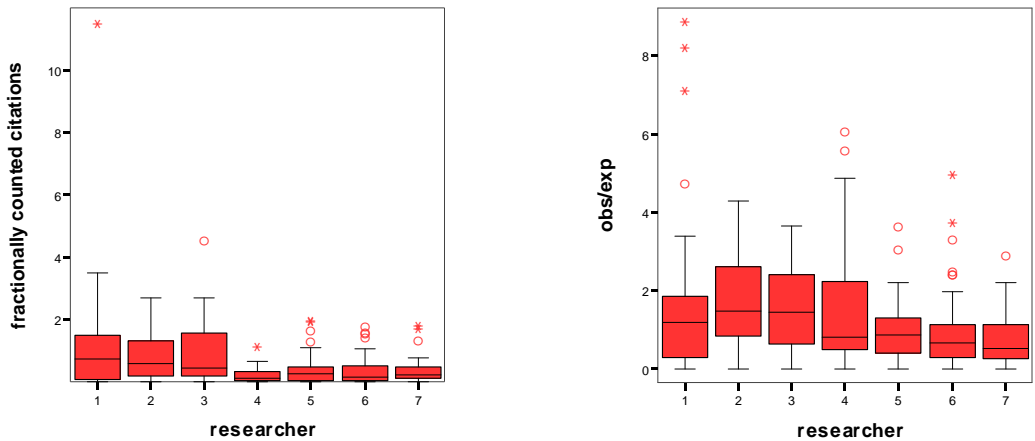

**Figure 2**: Boxplot of fractionally counted citations (left; this study) versus normalized citation rates (right; Opthof & Leydesdorff, 2010) for seven researchers in the AMC. (NB. Instead of means, medians are indicated as the lines in the box plots.)

Figure 2 shows box plots of the distributions of fractionated citations (left) compared to our previous results based on observed versus expected citation rates (right). Whereas we previously found, for example, no significant differences between the first and fourth authors in the case of journal normalization using a *post-hoc* test with Bonferroni correction, the first, second, and third author can be considered as a homogenous set on the basis of our field normalization. Furthermore, the third author's fractioned citation profile is significantly different from the fourth (using the Tukey test).[3] Using fractional counting for the normalization at the field level, one would be warranted to distinguish two groups among these seven researchers.

Note that by using fractional citation counts one abandons the notion of a world average as a standard for a field of science. Given the overlaps among fields, such a general standard is, in our opinion, sociologically unwarranted. By using fractional citation counts, however, one can benchmark against any reference set including the ones subsumed under the 221 ISI Subject Categories or the 60 subfields distinguished by ECOOM (Glänzel & Schubert, 2003; Glänzel *et al.* 2009). An additional advantage is that one can then use standard statistics to determine whether the performance above or below this "world average" is significant. A further extension to non-parametric statistics as advocated by Bornmann (2010; cf. Leydesdorff, 1990; Plomp, 1990) remains possible.

This measure of fractional counting can be generalized as normalization for any differences in citation behavior among citing authors (Small & Sweeney, 1985). The resulting distributions can be analyzed statistically; error bars consequently can be indicated in the graphical results. The importation of indexer-based and potentially biased schemes of classification is no longer necessary. In another context (Leydesdorff & Opthof, 2010), we show that this measure can also be used to normalize the impact of journals by considering the citable issues in the denominator of the ISI-Impact Factor as a document set (in the years $t-1$ and $t-2$) which can be counted fractionally in terms of citations in the year $t$ (in the numerator). Thus, the measure is very general. As noted, we consider the Bonferroni correction *ex post* and its further refinements (e.g., the Tukey and Scheffé tests) as appropriate for testing significance among different sets.[4] These tests are available in statistical packages such as SPSS.

---

[3] The Bonferroni correction is often considered as too conservative. The equivalent Tukey test in SPSS is corrected for multiple comparisons (in addition to dyadic ones).

**References**

Bensman, S. J., & Leydesdorff, L. (2009). Definition and Identification of Journals as Bibliographic and Subject Entities: Librarianship vs. ISI Journal Citation Reports (JCR) Methods and their Effect on Citation Measures. *Journal of the American Society for Information Science and Technology, 60*(6), 1097-1117.

Bornmann, L. (2010). Towards an ideal method of measuring research performance: Some comments to the Opthof and Leydesdorff (in press) paper. *Journal of Informetrics, 4*(3), in print.

CWTS (2008). AMC-specifieke CWTS-analyse 1997-2006. Leiden: CWTS [access via AMC intranet; unpublished, confidential].

Garfield, E. (1979). Is citation analysis a legitimate evaluation tool? *Scientometrics, 1*(4), 359-375.

Glänzel, W., & Schubert, A. (2003). A new classification scheme of science fields and subfields designed for scientometric evaluation purposes. *Scientometrics, 56*(3), 357-367.

Glänzel, W., Thijs, B., Schubert, A., & Debackere, K. (2009). Subfield-specific normalized relative indicators and a new generation of relational charts: Methodological foundations illustrated on the assessment of institutional research performance. *Scientometrics, 78*(1), 165-188.

Leydesdorff, L. (1990). Relations Among Science Indicators I. The Static Model,. *Scientometrics 18*, 281-307.

Leydesdorff, L., & Opthof, T. (2010). *Scopus'* Source Normalized Impact per Paper (*SNIP*) versus the Journal Impact Factor based on fractional counting of citations, *Journal of the American Society for Information Science & Technology* (in print).

Leydesdorff, L., & Opthof, T. (in preparation). Normalization, CWTS indicators, and the Leiden Rankings: Differences in citation behavior at the level of fields, available at http://arxiv.org/abs/1003.3977.

McAllister, P. R., Narin, F., & Corrigan, J. G. (1983). Programmatic Evaluation and Comparison Based on Standardized Citation Scores. *IEEE Transactions on Engineering Management, 30*(4), 205-211.

Moed, H. F. (2010a). CWTS crown indicator measures citation impact of a research group's publication oeuvre, *Journal of Informetrics* 4(3), in print.

Moed, H. F. (2010b). Measuring contextual citation impact of scientific journals. *Journal of Informetrics,* in print.

Opthof, T., & Leydesdorff, L. (2010). Caveats for the journal and field normalizations in the CWTS ("Leiden") evaluations of research performance. *Journal of Informetrics,* in print.

---

[4] Additionally, a robust test of the equality of the means is provided by the Welch statistics (Glänzel, *personal communication*, March 18, 2010). One can also use Kruskall-Wallis for this purpose.

Plomp, R. (1990). The significance of the number of highly cited papers as an indicator of scientific prolificacy. *Scientometrics, 19*(3), 185-197.

Pudovkin, A. I., & Garfield, E. (2002). Algorithmic procedure for finding semantically related journals. *Journal of the American Society for Information Science and Technology,* 53(13), 1113-1119.

Rafols, I., & Leydesdorff, L. (2009). Content-based and Algorithmic Classifications of Journals: Perspectives on the Dynamics of Scientific Communication and Indexer Effects, *Journal of the American Society for Information Science and Technology,* 60(9), 1823-1835.

Small, H., & Sweeney, E. (1985). Clustering the Science Citation Index Using Co-Citations I. A Comparison of Methods, *Scientometrics 7*, 391-409.

Van Raan, A. F. J., Van Leeuwen, T. N., Visser, M. S., Van Eck, N. J., & Waltman, L. (2010). Rivals for the crown: Reply to Opthof and Leydesdorff. *Journal of Informetrics,* 4(3), in print.

---

[i] Amsterdam School of Communications Research (ASCoR), University of Amsterdam, Kloveniersburgwal 48, 1012 CX Amsterdam, The Netherlands; loet@leydesdorff.net; http://www.leydesdorff.net.

[ii] Experimental Cardiology Group, Heart Failure Research Center, Academic Medical Center AMC, Meibergdreef 9, 1105 AZ Amsterdam, The Netherlands; Department of Medical Physiology, University Medical Center Utrecht, Utrecht, The Netherlands.